\documentclass[a4paper,aps,prd,10pt,preprintnumbers,showpacs,twocolumn,groupedaddress,superscriptaddress,nofootinbib,amsmath,amssymb]{revtex4-1}\usepackage{graphicx}\def\PRDstyle#1{#1}\def\JCAPstyle#1{}\def\Abstract#1{\begin{abstract}#1\end{abstract}}
\usepackage[utf8]{inputenc}
\usepackage[T1]{fontenc}
\usepackage{cmap}
\usepackage[normalem]{ulem}

\def\imo{i}

\def\Im#1{\mathrm{Im}(#1)}
\def\n{\tilde{n}}
\def\Order#1{{\cal O}\left(#1\right)}

\begin{document}

\title{Nonoscillatory gravitational quasinormal modes and telling tails for Schwarzschild-de Sitter black holes}

\JCAPstyle{
\author[1]{R.~A.~Konoplya,}\emailAdd{roman.konoplya@gmail.com}
\author[2]{A.~Zhidenko}\emailAdd{olexandr.zhydenko@ufabc.edu.br}
\affiliation[1]{Research Centre for Theoretical Physics and Astrophysics, \\ Institute of Physics, Silesian University in Opava, \\ Bezručovo náměstí 13, CZ-74601 Opava, Czech Republic}
\affiliation[2]{Centro de Matemática, Computação e Cognição (CMCC), \\ Universidade Federal do ABC (UFABC), \\ Rua Abolição, CEP: 09210-180, Santo André, SP, Brazil}
\arxivnumber{2209.12058}
}

\PRDstyle{
\author{R. A. Konoplya}\email{roman.konoplya@gmail.com}
\affiliation{Research Centre for Theoretical Physics and Astrophysics, Institute of Physics, Silesian University in Opava, Bezručovo náměstí 13, CZ-74601 Opava, Czech Republic}
\author{A. Zhidenko} \email{olexandr.zhydenko@ufabc.edu.br}
\affiliation{Centro de Matemática, Computação e Cognição (CMCC), Universidade Federal do ABC (UFABC), \\ Rua Abolição, CEP: 09210-180, Santo André, SP, Brazil}
}

\Abstract{
We show that the quasinormal spectrum of gravitational perturbations of Schwarzschild - de Sitter black holes contains a new branch of purely imaginary modes. These modes are not algebraically special and we showed that the sum of them form the well-known in the literature exponential asymptotic tail. When the ratio of the event horizon radius to the cosmological horizon vanishes, these quasinormal modes approach modes of empty de Sitter spacetime. Thus, the spectrum consists of the two branches: Schwarzschild branch deformed by the cosmological constant and de Sitter branch deformed by the black hole mass.
While the de Sitter branch contains purely imaginary modes only, the oscillatory modes (with nonzero real part) of the Schwarzschild branch can also become purely imaginary for some values of the cosmological constant, for which they approach the algebraically special mode.
}

\maketitle

\section{Introduction}\label{sec:Intro}

Quasinormal modes~\cite{reviews} are an essential characteristic of a black hole geometry, which is currently observed via gravitational waves interferometers~\cite{LIGOScientific:2016aoc,LIGOScientific:2017vwq,LIGOScientific:2020zkf}. The quasinormal spectrum of a number of classical solutions of the Einstein equations, such as Schwarzschild, Reissner-Nordström, and Kerr, have been exhaustively studied. An extensive investigation of the quasinormal spectrum of such black holes immersed in an asymptotically de Sitter world have been done for gravitational and test fields~\cite{Zhidenko:2003wq,Yoshida:2003zz,Konoplya:2004uk,Cardoso:2003sw,Tattersall:2018axd,Aragon:2020tvq,Hintz:2021vfl,Churilova:2021nnc,Roussev:2022hpm,Konoplya:2007zx,Dyatlov:2011jd} allowing also for an electric charge and nonzero rotation of the black hole. The literature on quasinormal modes of asymptotically de Sitter black holes in alternative and higher dimensional gravity is immense by now (see, for example~\cite{Konoplya:2003dd,Molina:2003ff,Fernando:2015fha,Natario:2004jd,Konoplya:2008au,Konoplya:2017ymp,Cuyubamba:2016cug,Konoplya:2013sba} and references therein).
Considerable interest to quasinormal modes of asymptotically de Sitter black holes have been recently paid due to the possible Strong Cosmic Censorship bound on quasinormal modes~\cite{Cardoso:2017soq,Dias:2018ynt,Dias:2018ufh,Hod:2020ktb,Liu:2019lon}.

It is well known that the quasinormal modes do not form a complete set~\cite{reviews}, so that the signal can be represented as a sum of quasinormal frequencies only at some intermediate stage of the decay, while at asymptotically late times, nonoscillatory tails dominate in a signal. Thus, quasinormal modes are usually clearly distinguished from asymptotic tails.
Relaxation of perturbations of asymptotically de Sitter spacetimes is also remarkable in the time domain in this aspect: instead of the usual power-law tail, appropriate to an asymptotically flat case~\cite{Price:1971fb}, {\it exponential} tails dominate at late times, as was shown in~\cite{Brady:1996za} for scalar field perturbations and~\cite{Brady:1999wd} for gravitational and other spin fields. Actually, the accurate law for asymptotic tails cannot be easily extracted from the numerical data because of the complex dependence on all the parameters~\cite{Molina:2003dc}. Although the exponential dependence of the tails could make one suspect that this is another form of the quasinormal stage, no such guess was spoken in the literature, to the best of our knowledge.

Looking at such extensive study of quasinormal spectra of asymptotically de Sitter black holes, one could not expect that there is an essential gap in our knowledge of gravitational perturbations of a simple four-dimensional Schwarzschild-de Sitter black hole. Here we will show that there is a new branch of gravitational quasinormal modes of the Schwarzschild-de Sitter spacetime with peculiar properties: the modes are purely imaginary, that is, nonoscillatory, and at asymptotically late times they form the telling tail found for the first time in~\cite{Brady:1996za}. Unlike the algebraically special mode, these purely imaginary modes satisfy the purely ingoing boundary condition at the event horizon and, therefore, can be considered as part of the quasinormal spectrum. When the mass of the black hole goes to zero, these modes approach modes of the empty de Sitter spacetime~\cite{Lopez-Ortega:2006aal}. When the cosmological constant is vanishing, the purely imaginary quasinormal modes approach zero, so that from the spectral point of view these modes have the least damping rate and dominate in the spectrum when the cosmological constant is small, as it is prescribed by the observational cosmology. Similar purely imaginary quasinormal modes were found for a test scalar field in~\cite{Jansen:2017oag,Cardoso:2017soq}, but, to the best of our knowledge, no analysis was done for the gravitational sector.

The paper is organized as follows: In Sec.~\ref{sec:wavelike} we consider the wavelike equation for the gravitational perturbations of Schwarzschild-de Sitter black hole. Sec.~\ref{sec:Frobenius} is devoted to the main numerical method which we used in the frequency domain: the Frobenius method used by Leaver for finding quasinormal modes~\cite{Leaver:1985ax}. In Sec.~\ref{sec:Degenerate} algebraically special modes are deduced from a degenerate case of the Frobenius method. Sec.~\ref{sec:Bernstein} is devoted to another method in frequency domain, the Bernstein spectral method, which we used in order to check our results obtained by the Frobenius method. The time-domain integration method is briefly discussed in Sec.~\ref{sec:timedomain}. In Sec.~\ref{sec:qnms} we discuss the properties of the new branch of quasinormal modes, while in Sec.~\ref{sec:tails} we study exponential asymptotic tail and show how they are related with the new branch of quasinormal modes. Finally, in Sec.~\ref{sec:conclusion} we summarize the obtained results and discuss a number of open questions.

\section{Wave equations for SdS black hole}\label{sec:wavelike}

The Schwarzschild-de-Sitter black hole is described by the metric
\begin{eqnarray}\label{SdS-metric}%
ds^2 &=& f(r)dt^2 - \frac{dr^2}{f(r)} - r^2 (d\theta^2 + \sin^2\theta d\phi^2),
\\\nonumber
f(r) &=& 1 - \frac{2M}{r} - \frac{\Lambda r^2}{3}
\JCAPstyle{=}
\PRDstyle{\\\nonumber&=&}
\Lambda\frac{(r_c-r)(r-r_e)(r+r_e+r_c)}{3r},
\end{eqnarray}
where
$$M=\frac{r_c r_e (r_c + r_e)}{2 (r_c^2 + r_c r_e + r_e^2)}$$
is the black hole mass, and
$$\Lambda=\frac{3}{r_c^2 + r_c r_e + r_e^2}$$
is the cosmological constant. The quantities $r_e$ and $r_c$ are the radii of the event and cosmological horizons respectively.

It is well known that after some algebra the perturbation equations of the Schwarzschild-de Sitter spacetime can be reduced to the Schrödinger wavelike equation
\begin{equation}\label{Wave-like-equation}%
\left(\frac{d^2}{dr_*^2} + \omega^2 - V(r_*)\right)\Phi(r_*) = 0,
\end{equation}
with respect to the tortoise coordinate,
\begin{equation}\label{tortoise-coordinate}
dr_* \equiv \frac{dr}{f(r)}.
\end{equation}

Under the choice of the positive sign of the real part of $\omega$, QNMs satisfy the following boundary conditions
\begin{equation}\label{bounds}
\Phi(r_*) \propto e^{\pm\imo\omega r_*}, \qquad r_*\to\pm\infty,
\end{equation}
corresponding to purely in-going waves at the event horizon and purely out-going waves at the cosmological horizon.

The effective potential for the axial gravitational perturbations is given by the following expression (see, for instance,~\cite{Moss:2001ga}):
\begin{equation}\label{potential}
V(r) = f(r)\left(\frac{\ell(\ell+1)}{r^2} - \frac{6M}{r^3}\right),
\end{equation}
where $\ell=2,3,4,\ldots$ is the multipole number. It is analytically proved and well known that the polar perturbations are isospectral with the axial ones, so that only one type of perturbations is sufficient for our consideration.

\section{Frobenius method}\label{sec:Frobenius}

The main numerical method which we will use in the frequency domain for finding accurate values of quasinormal modes is based on the Frobenius series expansion. It was applied by Leaver~\cite{Leaver:1985ax} for a black hole spectral problem for the first time. Having in mind that this method is well-known we will only briefly discuss it here, referring a reader to \cite{reviews} for more details.

The wavelike equation (\ref{Wave-like-equation}) has three regular singular points: $r=r_e$, $r=r_c$, and $r=-r_e+r_c$. The appropriate Frobenius series are \cite{Yoshida:2003zz}
\begin{eqnarray}
\Phi(r_*)&=&\left(\frac{1}{r_e} - \frac{1}{r}\right)^{\rho_e}\left(\frac{1}{r}-\frac{1}{r_c}\right)^{-\rho_c}\left(\frac{1}{r}+\frac{1}{r_e+r_c}\right)^{\rho_c+\rho_e}
\PRDstyle{\nonumber\\&&\times}\sum_{n=0}^{\infty}a_n\left(\frac{\frac{1}{r}-\frac{1}{r_e}}{\frac{1}{r_c}-\frac{1}{r_e}}\right)^n,
\label{Frobenius}
\end{eqnarray}
where
$r_e$ is the event horizon, $r_c$ is the cosmological horizon, and $\rho_e$ and $\rho_c$ are given by
$$e^{\imo\omega r_*}=\left(\frac{1}{r} - \frac{1}{r_e}\right)^{-\rho_e}\left(\frac{1}{r_c}-\frac{1}r\right)^{-\rho_c}\left(\frac{1}{r}+\frac{1}{r_e+r_c}\right)^{\rho_c+\rho_e}\!\!\!\!.$$
Then, one can find that
\begin{eqnarray}\nonumber
&&\rho_e=\frac{\imo\omega}{\displaystyle 2M\left(\frac{1}{r_c}-\frac{1}{r_e}\right)\left(\frac{1}{r_c+r_e}+\frac{1}{r_e}\right)},
\JCAPstyle{\qquad}\PRDstyle{\\\nonumber&&}
\rho_c=\frac{\imo\omega}{\displaystyle 2M\left(\frac{1}{r_e}-\frac{1}{r_c}\right)\left(\frac{1}{r_c+r_e}+\frac{1}{r_c}\right)}.
\end{eqnarray}

Substituting (\ref{Frobenius}) into (\ref{Wave-like-equation}), we obtain the three-terms recurrent relation for $a_n$
\begin{equation}\label{reccur}
a_{n+1}\alpha_n+a_n\beta_n+a_{n-1}\gamma_n=0, \qquad n\geq0, \quad \gamma_0=0,
\end{equation}
where the coefficients $\alpha_n, \beta_n, \gamma_n$ have the form \cite{Yoshida:2003zz}
\begin{eqnarray}\nonumber
\alpha_n&=&\dfrac{r_c(r_c + 2 r_e) (1 + n) (1 + n + 2 \rho_e)} {r_c^2 + r_c r_e + r_e^2},
\\\nonumber
\beta_n&=&-\dfrac{(n+2\rho_e)(n+2\rho_e+1)(2 r_c^2 + 2 r_c r_e - r_e^2)}{r_c^2 + r_c r_e + r_e^2}
\PRDstyle{\\\nonumber&&}
- \ell(\ell+1) + 3\dfrac{r_c(r_c+r_e)}{r_c^2 + r_c r_e + r_e^2},
\\
\gamma_n&=&\dfrac{r_c^2 -r_e^2 }{r_c^2 + r_c r_e + r_e^2}((n+2\rho_e)^2-4).
\label{coefficients}
\end{eqnarray}

Following Leaver~\cite{Leaver:1985ax}, we are searching QNMs as the most stable roots of the following algebraic equation:
\begin{eqnarray}\label{continued_fraction}
\beta_n&-&\frac{\alpha_{n-1}\gamma_{n}}{\beta_{n-1}
-\frac{\alpha_{n-2}\gamma_{n-1}}{\beta_{n-2}-\frac{\alpha_{n-3}\gamma_{n-2}}{\beta_{n-3}-\ldots}}}
\PRDstyle{\\\nonumber&&}
=\frac{\alpha_n\gamma_{n+1}}{\beta_{n+1}-\frac{\alpha_{n+1}\gamma_{n+2}}{\beta_{n+2}-\frac{\alpha_{n+2}\gamma_{n+3}}{\beta_{n+3}-\ldots}}}.
\end{eqnarray}
We employ the Nollert method~\cite{Nollert:1993zz} in order to improve convergence of the right-hand side of equation (\ref{continued_fraction}) and, thereby, to compute higher overtones in a quicker way.

\section{Degenerate case}\label{sec:Degenerate}

A particular degenerate case of the Frobenius series must be considered separately. It corresponds to a purely imaginary frequency, for which $N + 2 \rho_e=0$ for some integer $N$, implying that $\alpha_{N-1}=\gamma_{N-2}=0$.
In this case Eq.~(\ref{reccur}) reads
\begin{equation}\label{degenerate}
\begin{array}{rcl}
a_{N-1}\beta_{N-1}+a_{N-2}\gamma_{N-1}&=&0,\\
a_{N-1}\alpha_{N-2}+a_{N-2}\beta_{N-2}&=&0,
\end{array}
\end{equation}
and all other coefficients are zero,
$$a_0=a_1=\ldots=a_{N-3}=a_{N}=a_{N-1}=\ldots=0.$$
The consistency condition is
\begin{equation}\label{consistency}
-\frac{a_{N-1}}{a_{N-2}}=\frac{\gamma_{N-1}}{\beta_{N-1}}=\frac{\beta_{N-2}}{\alpha_{N-2}},
\end{equation}
which is equivalent to the relation
\begin{equation}\label{spec}
N=\frac{(\ell-1)\ell(\ell+1)(\ell+2)(r_c^2+r_cr_e+r_e^2)^2}{3r_c(r_c-r_e)(r_c+r_e)(r_c+2r_e)}.
\end{equation}
This corresponds to the well-known algebraically special frequency, which was first found for the asymptotically flat Schwarzschild black hole \cite{Chandrasekhar}:
\begin{equation}
\omega_a=-\imo\frac{(\ell-1)\ell(\ell+1)(\ell+2)}{12M}.
\end{equation}
Although this frequency satisfies equation (\ref{continued_fraction}), which is reduced to (\ref{consistency}) in this case, it is clear that it does not satisfy the quasinormal boundary condition (ingoing wave) at the horizon.
Indeed, since $a_0=a_1=\ldots=a_{n-3}=0$, for $r\to r_e$ ($r_*\to-\infty$)
$$\Phi\propto(r-r_e)^{\rho_e+N-2}=(r-r_e)^{-\rho_e-2}\propto e^{\imo\omega_a r_*},$$
which corresponds to the outgoing wave.

It is possible to find a general solution to the wavelike equation (\ref{Wave-like-equation}), for $\omega=\omega_a$.
The reason is that the effective potential (\ref{potential}) obeys the following relation:
\begin{eqnarray}\label{potalgebraic}
&&V(r_*)=A(r_*)^2+2\imo\omega_aA(r_*)-\frac{dA}{dr_*},
\JCAPstyle{\qquad}\PRDstyle{\\\nonumber&&}
A(r)=f(r)\frac{6M}{6M r+(\ell-1)(\ell+2)r^2},
\end{eqnarray}
and consequently for $\omega=\omega_a$, eq.~(\ref{Wave-like-equation}) can be rewritten as
\begin{equation}\label{Riccati}
\frac{1}{\Phi(r_*)}\frac{d^2\Phi}{dr_*^2}=(A(r_*)+\imo\omega_a)^2-\frac{dA}{dr_*},
\end{equation}
The general solution to Eq.~(\ref{Riccati}) is
\begin{eqnarray}\label{gensol}
&&\Phi=C_i\left((\ell-1)(\ell+2) + \frac{6 M}{r}\right)e^{-\imo\omega_a r_*}+
\\\nonumber&&
\JCAPstyle{\qquad}C_o\left((\ell-1)(\ell+2)+2\imo \omega_a r_e + \frac{6M-\ell(\ell+1)r_e}{r}\right)e^{\imo\omega_a r_*}.
\end{eqnarray}
The degenerate case of the Frobenius series which we consider corresponds to $C_i=0$.

For $M>0$ this solution does not obey the quasinormal boundary conditions (\ref{bounds}), however, for $M<0$, it corresponds to the bound state with $\Im{\omega_a}>0$, which governs the instability \cite{Cardoso:2006bv}.

From the above consideration it is possible to conclude that the formal usage of the Frobenius method provides also a nonquasinormal solution (algebraically special mode which does not satisfy the quasinormal boundary conditions) for such values of $\ell$, $r_e$, and $r_c$, that the right-hand side of (\ref{spec}) is integer. However, since the solution to the equation (\ref{continued_fraction}) are continuous functions of the parameters, one can expect that there are modes in the Schwarzschild-de Sitter black hole spectrum, which satisfy the quasinormal boundary conditions and approach the algebraically special mode when (\ref{spec}) becomes an integer. In order to separate numerically these modes from the algebraically special one, we solve Eq.~(\ref{continued_fraction}) for such values of $r_e/r_c$, that $N$ in (\ref{spec}) is not integer.

\begin{figure*}
\resizebox{\linewidth}{!}{\includegraphics{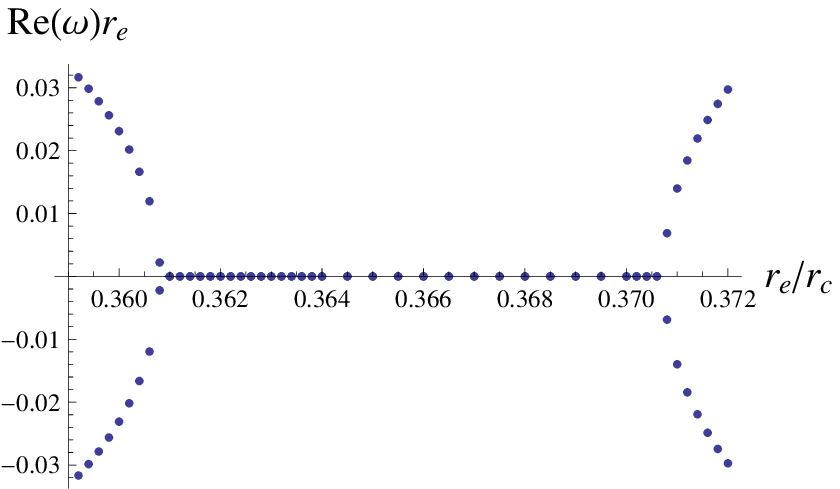}\includegraphics{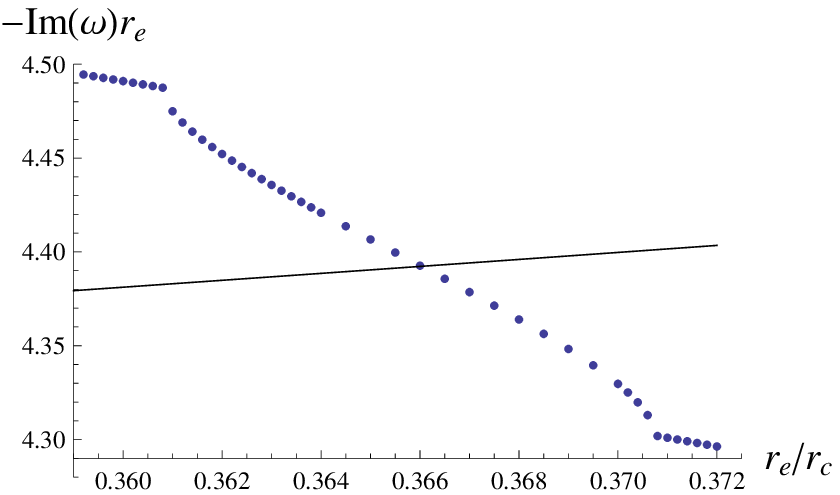}}
\caption{Algebraically special mode for $\ell=2$ (black solid), and the numerical values (real and imaginary parts) for the quasinormal mode, which approaches the degenerate case at $r_e/r_c=(\sqrt{3}-1)/2\approx0.366$ (in (\ref{spec}) it corresponds to $N=12$). }\label{fig:algebraic}
\end{figure*}

Fig.~\ref{fig:algebraic} illustrates the behavior of the mode, which becomes algebraically special at $r_e/r_c=(\sqrt{3}-1)/2$, corresponding to $N=12$. One can see that it becomes purely imaginary in the parametric region near the point where it approaches the algebraically special mode. Otherwise the mode has a nonzero real part and, therefore, it does not correspond to the purely imaginary quasinormal modes of the empty de Sitter spacetime. It is located between the two modes presented in (\ref{fig:dominant}).

\section{Bernstein spectral method}\label{sec:Bernstein}

While the Frobenius method is quickly converging, it is possible simply to miss this or that mode while searching the roots of the equation with the infinite continued fraction.
In order to check it, we can use the Bernstein spectral method which is especially good for detecting purely imaginary frequencies, though its computational complexity grows fast for higher overtones. That is why here we use it as a complementary method.

Following \cite{Fortuna:2020obg}, we introduce the function $\phi(u)$, which is regular for $0\leq u\leq 1$ when $\omega$ is a quasinormal mode,
\begin{equation}\label{regularized}
\Phi(r_*)=\left(\frac{1}{r_e} - \frac{1}{r}\right)^{\rho_e}\left(\frac{1}{r}-\frac{1}{r_c}\right)^{-\rho_c}\phi\left(\frac{\frac{1}{r}-\frac{1}{r_e}}{\frac{1}{r_c}-\frac{1}{r_e}}\right),
\end{equation}
and the compact coordinate $u$ is defined as follows:
$$u\equiv\frac{\frac{1}{r}-\frac{1}{r_e}}{\frac{1}{r_c}-\frac{1}{r_e}}.$$
We represent $\phi(u)$ as a sum
\begin{equation}\label{Bernsteinsum}
\phi(u)=\sum_{k=0}^NC_kB_k^N(u),
\end{equation}
where
$$B_k^N(u)\equiv\frac{N!}{k!(N-k)!)}u^k(1-u)^{N-k}$$
are the Bernstein polynomials.

Substituting (\ref{regularized}) into (\ref{Wave-like-equation}) and using a Chebyschev collocation grid of $N+1$ points,
$$u_p=\frac{1-\cos \frac{p\cdot\pi}{N}}{2}=\sin^2\frac{p\cdot\pi}{2N}, \qquad p=\overline{0,N},$$
we obtain a set of linear equations with respect to $C_k$, which has nontrivial solutions iff the corresponding coefficient matrix is singular. Since the elements of the coefficient matrix are polynomials (of degree 2) of $\omega$, the problem is reduced to the eigenvalue problem of a matrix pencil (of order 2) with respect to $\omega$, which can be solved numerically. Once the eigenvalue problem is solved, one can calculate the corresponding coefficients $C_k$ and explicitly determine the polynomial (\ref{Bernsteinsum}), which approximates the solution to the wave equation (\ref{Wave-like-equation}).

In order to exclude the spurious eigenvalues, which appear due to finiteness of the polynomial basis in (\ref{Bernsteinsum}), we compare both the eigenfrequencies and corresponding approximating polynomials for different values of $N$. Namely, for the coinciding eigenfrequencies, $\omega^{(1)}$ and $\omega^{(2)}$, obtained, respectively, for $N=N^{(1)}$ and $N=N^{(2)}$, we calculate
$$1-\frac{|\langle \phi^{(1)}\;|\;\phi^{(2)} \rangle|^2}{||\phi^{(1)}||^2||\phi^{(2)}||^2}=\sin^2\alpha,$$
where $\alpha$ is the angle between the vectors $\phi^{(1)}$ and $\phi^{(2)}$ in the $L^2$-space\footnote{Notice that the solution (eigenfunction) is defined up to an arbitrary constant factor. It was proposed in \cite{Fortuna:2020obg} to compare the normalized polynomials, e.~g., such that $\phi(0)=1$. However, fixing the value of polynomials in some point leads to additional numerical errors. That is why we compare the obtained polynomial approximations without a normalization.}. If $\alpha$ is sufficiently small, the difference between $\omega^{(1)}$ and $\omega^{(2)}$ provides the error estimation\footnote{The Wolfram Mathematica\textregistered{} package with the implementation of the Bernstein spectral method is publicly available from \url{https://arxiv.org/src/2211.02997/anc}.}.

The method allows one to determine the dominant quasinormal frequencies and the purely imaginary modes. The procedure converges faster at the purely imaginary modes. Typically, by using $N=50$, we obtained from $6$ to $14$ correct decimal places.

\section{Time-domain integration}\label{sec:timedomain}

In order to understand the role of newly found purely imaginary modes in the asymptotic late-time relaxation of perturbations we will use the time-domain integration of the wave equation at a fixed value of the radial coordinate. We integrate the wavelike equation in terms of the light-cone variables $u=t-r_*$ and $v=t+r_*$ via applying the discretization scheme of Gundlach-Price-Pullin \cite{Gundlach:1993tp},
\begin{eqnarray}\label{Discretization}
\Psi\left(N\right)&=&\Psi\left(W\right)+\Psi\left(E\right)-\Psi\left(S\right)\\\nonumber&&
-\Delta^2V\left(S\right)\frac{\Psi\left(W\right)+\Psi\left(E\right)}{4}+{\cal O}\left(\Delta^4\right)\,,
\end{eqnarray}
where the following notation for the points was used:
$N\equiv\left(u+\Delta,v+\Delta\right)$, $W\equiv\left(u+\Delta,v\right)$, $E\equiv\left(u,v+\Delta\right)$, and $S\equiv\left(u,v\right)$. The Gaussian initial data are imposed on the two null surfaces, $u=u_0$ and $v=v_0$. The dominant quasinormal frequencies can be extracted from the time-domain profiles with the help of the Prony method of fitting of the profile data by superposition of damped exponents,
\begin{equation}\label{Pronyfitexp}
\Psi(t)\simeq\sum_{k=1}^p C_ke^{-\imo\omega_k t},
\end{equation}
see, e.~g.,~\cite{Konoplya:2011qq}.

\section{Quasinormal modes}\label{sec:qnms}

\begin{table*}
\begin{tabular}{l@{\hspace{1em}}|@{\hspace{1em}}l@{\hspace{1em}}l@{\hspace{1em}}l@{\hspace{1em}}l}
\hline
$r_e/r_c$ & $\omega$ ($\n=1$) & $\omega$ ($\n=2$) & $\omega$ ($\n=3$) & $\omega$ ($\n=4$) \\
\hline
$0.05$ & $-0.14628837140338 \imo$ & $-0.19522273397247 \imo$ & $-0.2442623723715119 \imo$ & $-0.2934098923986 \imo$ \\
$0.07$ & $-0.20275478960143 \imo$ & $-0.27077350507009 \imo$ & $-0.3390767925979484 \imo$ & $-0.4076472923256 \imo$ \\
$0.08$ & $-0.23055584739888 \imo$ & $-0.30803805454118 \imo$ & $-0.3859275326664819 \imo$ & $-0.4641934521945 \imo$ \\
$0.1 $ & $-0.28530126566124 \imo$ & $-0.38156411560595 \imo$ & $-0.4785536394272280 \imo$ & $-0.5761938451731 \imo$ \\
$0.12$ & $-0.33891366229288 \imo$ & $-0.45377468788892 \imo$ & $-0.5697763064973273 \imo$ & $-0.6867708701612 \imo$ \\
$0.15$ & $-0.41722954207182 \imo$ & $-0.55965706830091 \imo$ & $-0.7040074373138981 \imo$ & $-0.8499582256716 \imo$ \\
$0.17$ & $-0.46805136531524 \imo$ & $-0.62863285222395 \imo$ & $-0.7917483098715319 \imo$ & $-0.9568774748040 \imo$ \\
$0.2 $ & $-0.54222246032764 \imo$ & $-0.72968015874795 \imo$ & $-0.9206496444050022 \imo$ & $-1.1143355632697 \imo$ \\
$0.25$ & $-0.66042067251270 \imo$ & $-0.89156948641743 \imo$ & $-1.1280908778179555 \imo$ & $-1.3684205445489 \imo$ \\
$0.30$ & $-0.77196917103691 \imo$ & $-1.04528324956367 \imo$ & $-1.3257123954840378 \imo$ & $-1.6115337897905 \imo$ \\
$0.35$ & $-0.87703503386909 \imo$ & $-1.19069916619331 \imo$ & $-1.5135744221049816 \imo$ & $-1.8427892217186 \imo$ \\
$0.40$ & $-0.97588626301967 \imo$ & $-1.32811425963975 \imo$ & $-1.6912980603515071 \imo$ & $-2.0617237836981 \imo$ \\
$0.50$ & $-1.15577483691395 \imo$ & $-1.57868577827844 \imo$ & $-2.0160503960899165 \imo$ & $-2.4612386253633 \imo$ \\
$0.60$ & $-1.31399827401026 \imo$ & $-1.79969050137711 \imo$ & $-2.3014203318274638 \imo$ & $-2.8127817053707 \imo$ \\
$0.70$ & $-1.45287575152641 \imo$ & $-1.99316218920237 \imo$ & $-2.5517982132368326 \imo$ & $-3.1256949854711 \imo$ \\
$0.80$ & $-1.57482840046419 \imo$ & $-2.16235944515726 \imo$ & $-2.7728491352788873 \imo$ & $-3.39607         \imo$ \\
$0.90$ & $-1.68197827112948 \imo$ & \\
\hline
\end{tabular}
\caption{First four purely imaginary quasinormal modes found by the Bernstein polynomial method for $\ell=2$, $r_e=1$ and various values of $r_e/r_c$. }\label{table:rho}
\end{table*}

\begin{figure}
\resizebox{\linewidth}{!}{\includegraphics{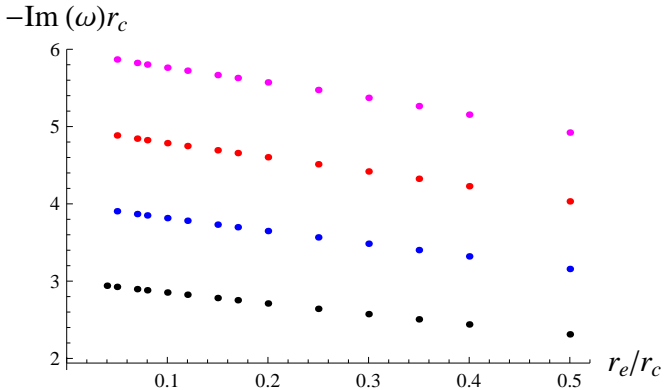}}
\caption{The first four purely imaginary mode as a function of $r_e/r_c$ for $\ell=2$. The accurate values of the modes are given in table~\ref{table:rho005}.}\label{fig:dominant}
\end{figure}

The gravitational quasinormal modes of the four dimensional empty de Sitter spacetime is \cite{Lopez-Ortega:2006aal}
\begin{equation}
\omega r_c = -\imo(\ell+\n),
\end{equation}
where $\n=1,2,\ldots$.
From Fig.~\ref{fig:dominant} we can see that the purely imaginary quasinormal modes can be very well fit by the following linear formula:
\begin{equation}\label{largerclimit}
\omega = -\imo\frac{\ell+\n}{r_c}\left(1-\frac{r_e}{2r_c} + \Order{\frac{r_e}{r_c}}^2\right).
\end{equation}
Notice that $\n$ numbers the modes of the purely imaginary branch and is, therefore, not the overtone number of the whole spectrum. Thus, the purely imaginary modes evidently go over into the quasinormal modes of empty de Sitter spacetime when the radius of the event horizon goes to zero.

From the above formula (\ref{largerclimit}) and table~\ref{table:rho} we can see that the damping rate of this branch of frequencies diminishes when the cosmological horizon $r_c$ is increased, so that they dominate at late times when $r_c$ is large. This way even tiny value of the cosmological constant drastically changes the properties of the quasinormal spectrum. A similar property, sensitivity of overtones to tiny changes solely near the event horizon, has been recently observed in \cite{Konoplya:2022hll,Konoplya:2022pbc}. Thus, here we may have another manifestation of sensitivity of the overtones to small changes of the geometry at the other boundary: near the cosmological horizon.

Summarizing this observation with earlier data on quasinormal modes of Schwarzschild-de Sitter black holes \cite{Zhidenko:2003wq,Konoplya:2004uk} we conclude that the quasinormal spectrum consists from the two qualitatively different parts:
\begin{itemize}
\item {\it Schwarzschild-like} quasinormal modes deformed by a nonzero value of the cosmological constant $\Lambda$. These modes go their Schwarzschild values when $\Lambda$ vanishes.
\item {\it de Sitter-like} quasinormal modes which are purely imaginary modes representing the spectrum of the empty de Sitter spacetime deformed by a black hole mass.
\end{itemize}

Since all the quasinormal modes of Schwarzschild-de Sitter black holes vanish in the extreme limit~\cite{Cardoso:2003sw}, it is clear that, as $\Lambda M^2$ (or, alternatively, $r_e/r_c$) grows, all the Schwarzschild-de Sitter quasinormal modes with the decay rate larger than the algebraically special mode, cross this mode at some point (becoming a purely imaginary mode in its vicinity) what can be seen in Fig.~\ref{fig:algebraic}. Thus, the algebraically special mode which does not satisfy the quasinormal boundary condition at the event horizon and does not depend on $\Lambda$ is surrounded by the purely imaginary quasinormal modes. Apparently, the latter ones go over into the algebraically special at particular values of $r_{e}/r_{c}$, for which $N$ in (\ref{spec}) is integer.

\begin{table}
\begin{tabular}{r@{\hspace{1em}}|@{\hspace{1em}}@{\hspace{1em}}l@{\hspace{1em}}l}
\hline
$\n$ & Bernstein & Frobenius \\
\hline
$ 1$ & $-0.1462883714034 \imo$ & $-0.1462883258500359 \imo$ \\
$ 2$ & $-0.1952227339725 \imo$ & $-0.1952173306184124 \imo$ \\
$ 3$ & $-0.2442623723715 \imo$ & $-0.2442582762738253 \imo$ \\
$ 4$ & $-0.2934098923986 \imo$ & $-0.2934072896458776 \imo$ \\
$ 5$ & $-0.3427          \imo$ & $-0.3426598512634874 \imo$ \\
$ 6$ &                         & $-0.3920109676819801 \imo$ \\
$ 7$ &                         & $-0.4414553386603660 \imo$ \\
$ 8$ &                         & $-0.4909875272073938 \imo$ \\
$ 9$ &                         & $-0.5406021350570267 \imo$ \\
$10$ &                         & $-0.5902939662635239 \imo$ \\
$11$ &                         & $-0.6400581387244684 \imo$ \\
$12$ &                         & $-0.6898900965406483 \imo$ \\
$13$ &                         & $-0.7397855009394569 \imo$ \\
$14$ &                         & $-0.7897400341227782 \imo$ \\
$15$ &                         & $-0.8397492168019296 \imo$ \\
\hline
\end{tabular}
\caption{Imaginary quasinormal modes found by the Berntein polynomial method and Frobenius method for $\ell=2$, $r_e=0.05r_c=1$. }\label{table:rho005}
\end{table}

\begin{table}
\begin{tabular}{c@{\hspace{1em}}|@{\hspace{1em}}@{\hspace{1em}}l@{\hspace{1em}}l}
\hline
$\n$ & Bernstein & Frobenius \\
\hline
$1$ & $-1.57482840046419 \imo$ & $-1.5748284004641904 \imo$ \\
$2$ & $-2.16235944515726 \imo$ & $-2.1623594451572610 \imo$ \\
$3$ & $-2.77284913527889 \imo$ & $-2.7728491352354840 \imo$ \\
$4$ & $-3.39607          \imo$ & $-3.3959708601254315 \imo$ \\
$5$ &                          & $-4.0200485275451892 \imo$ \\
$6$ &                          & $-4.6262009532229878 \imo$ \\
$7$ &                          & $-5.2441265168687816 \imo$ \\
$8$ &                          & $-5.9596491777120556 \imo$ \\
\hline
\end{tabular}
\caption{Imaginary quasinormal modes found by the Bernstein polynomial method and Frobenius method for $\ell=2$, $r_e=0.8r_c=1$. }\label{table:rho08}
\end{table}

In tables~\ref{table:rho005} and~\ref{table:rho08} we can see the numerical data for the purely imaginary quasinormal modes obtained with the Frobenius method and checked for a few first overtones by the Bernstein spectral method. Unfortunately, the latter approach converges too slowly at higher overtones and can be used mainly for detecting the correct spacing between the modes, so that one can be sure at the end of the day that no modes were missed by the Frobenius method.

\begin{table*}
\begin{tabular}{c@{\hspace{1em}}|@{\hspace{1em}}l@{\hspace{1em}}l@{\hspace{1em}}@{\hspace{1em}}l@{\hspace{1em}}l}
\hline
$\ell$ & $\omega$ ($\n=1$) & $\omega$ ($\n=2$) & $\omega$ ($\n=3$) & $\omega$ ($\n=4$) \\
\hline
$2$ & $-0.28530126566124 \imo$ & $-0.38156411560595 \imo$ & $-0.4785536394272280 \imo$ & $-0.5761938451731 \imo$ \\
$3$ & $-0.37992030383120 \imo$ & $-0.47558202160404 \imo$ & $-0.5717267881951159 \imo$ & $-0.6683323952569 \imo$ \\
$4$ & $-0.47473614477645 \imo$ & $-0.57016307245355 \imo$ & $-0.6659543089866820 \imo$ & $-0.7621195207195 \imo$ \\
$5$ & $-0.56960489303954 \imo$ & $-0.66490868869968 \imo$ & $-0.76051274255394   \imo$ & $-0.8565          \imo$ \\
$6$ & $-0.66449417925028 \imo$ & $-0.75972384841616 \imo$ & $-0.8552445897488    \imo$ & $-0.951           \imo$ \\
$7$ & $-0.75939340576028 \imo$ & $-0.854581         \imo$ & $-0.9502             \imo$ &                         \\
\hline
\end{tabular}
\caption{First four purely imaginary quasinormal modes found by the Bernstein polynomial method for $r_e=0.1r_c=1$, and various values of $\ell$. }\label{table:rho01}
\end{table*}

From table~\ref{table:rho01} we see that the purely imaginary modes have the same spacing for higher $\ell$.

\begin{figure}
\resizebox{1.01 \linewidth}{!}{\includegraphics{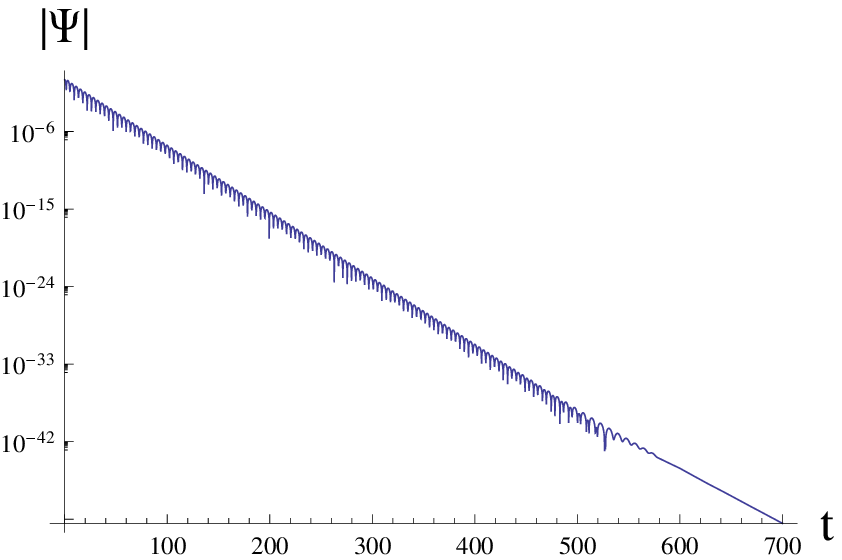}}
\caption{The time-domain profile for $\ell=2$, $r_e=0.05r_c=1$. At late time the Prony method gives the following dominant modes: $\omega_0=-0.14628832\imo$, $\omega_1=\pm0.743134-0.177149\imo$, and $\omega_2=-0.19520\imo$.}\label{fig:rho005}
\end{figure}

\section{Exponential late time tails}\label{sec:tails}

In 1996 P.~Brady, C~M.~Chambers, W.~Krivan, and P.~Laguna \cite{Brady:1996za} have shown that the decay of the massless scalar field in the Schwarzschild-de Sitter and Reissner-Nordström-de Sitter background are exponential at asymptotically late times. This behavior is different from the power-law tails of asymptotically flat black holes. In particular they showed that the scalar field falls off as
\begin{equation}\label{Brady}
\Psi \sim e^{-\ell k_{c} t},
\end{equation}
where $k_{c}$ is the surface gravity at the cosmological horizon. This work was further extended in~\cite{Brady:1999wd} and, finally generalized by C.~Molina and coworkers in~\cite{Molina:2003dc}, where it was shown that, strictly speaking, the above law (\ref{Brady}) is only approximate and the general exponential decay law depends on the spin of perturbations and the surface gravity at the cosmological horizon in a complex way
\begin{equation}
\Psi \sim e^{- \alpha (\ell, s, k_{c}) t},
\end{equation}
so that only some approximate values for constants $\alpha (\ell, s, k_{c})$ were provided in~\cite{Molina:2003dc}.
Thus, the accurate analytical formula for the asymptotic late time tails were unknown.

In Fig.~\ref{fig:rho005} we can see the time-domain profile for $\ell=2$ gravitational perturbations at $r_e=0.05r_c=1$. At late time, when the exponential tails dominate, the Prony method allows one to extract the following three dominant modes:
\begin{equation}\label{threedominant}
\begin{array}{rcl}
\omega_0&=&-0.14628832\imo,
\\
\omega_1&=&\pm0.743134-0.177149\imo,
\\
\omega_2&=&-0.19520\imo.
\end{array}
\end{equation}
Here we notice that $\omega_1$ is Schwarzschild-like mode deformed by the nonzero cosmological constant, while $\omega_0$ and $\omega_2$ are purely imaginary modes representing deformation of the de Sitter spacetime by a black hole mass. Thus, apparently, the exponential tail consists from contributions of all the purely imaginary quasinormal modes. That is the reason why it would be difficult to find a fit representing a general accurate analytical formula for the exponential tails.

\section{Conclusion}\label{sec:conclusion}

In the present paper we have found a new branch of quasinormal modes for gravitational perturbations of Schwazrschild-de Sitter spacetime. This modes possess a number of interesting and qualitatively different (from the asymptotically flat case) properties:
\begin{itemize}
\item Purely imaginary quasinormal modes approach quasinormal modes of the empty de Sitter spacetime, when the mass of the black hole goes to zero.
\item These modes vanish when the cosmological constant $\Lambda$ goes to zero, so that they become dominant at late times, once $\Lambda$ is small.
\item These modes form the well-known exponential ``telling tail'' at late times, which means that the asymptotic tail is simply a continuation of quasinormal decay at purely imaginary (nonoscillatory) modes.
\item The Schwarzschild-like branch of modes also contains purely imaginary quasinormal modes in a narrow range of values of the cosmological constant, so that they go over into the algebraically special mode. The latter does not satisfy the purely incoming wave boundary condition at the event horizon and is not quasinormal therefore.
\end{itemize}

Our paper could be extended to the case of nonzero electric charge and angular momentum (Reissner-Nordström-de Sitter and Kerr-Newman-de Sitter spacetime). A similar connection between the charged scalar field asymptotic tail and the known purely imaginary quasinormal modes for this field is expected, but, seems, has not be studied so far. An extension to asymptotically de Sitter black holes in higher dimensional and alternative theories of gravity could also be interesting in this context and we hope that future publications will shed light on all these questions.

\begin{acknowledgments}
A.~Z. was supported by Conselho Nacional de Desenvolvimento Científico e Tecnológico (CNPq).
R.~K. would like to acknowledge support of the grant No.~19-03950S of Czech Science Foundation (GAČR).
\end{acknowledgments}


\begin{thebibliography}{80}
\bibitem{reviews}
R.~A.~Konoplya and A.~Zhidenko,
Rev.\ Mod.\ Phys.\  {\bf 83}, 793 (2011)
[arXiv:1102.4014 [gr-qc]];\\
E.~Berti, V.~Cardoso and A.~O.~Starinets,
Class.\ Quant.\ Grav.\  {\bf 26}, 163001 (2009)
[arXiv:0905.2975 [gr-qc]];\\
K.~D.~Kokkotas and B.~G.~Schmidt,
Living Rev.\ Rel.\  {\bf 2}, 2 (1999)
[gr-qc/9909058];\\
H.~P.~Nollert,
Class.\ Quant.\ Grav.\  {\bf 16}, R159 (1999).

\bibitem{LIGOScientific:2016aoc}
B.~P.~Abbott \textit{et al.} [LIGO Scientific and Virgo],
Phys. Rev. Lett. \textbf{116}, no.6, 061102 (2016)
[arXiv:1602.03837 [gr-qc]].

\bibitem{LIGOScientific:2017vwq}
B.~P.~Abbott \textit{et al.} [LIGO Scientific and Virgo],
Phys. Rev. Lett. \textbf{119}, no.16, 161101 (2017)
[arXiv:1710.05832 [gr-qc]].

\bibitem{LIGOScientific:2020zkf}
R.~Abbott \textit{et al.} [LIGO Scientific and Virgo],
Astrophys. J. Lett. \textbf{896}, no.2, L44 (2020)
[arXiv:2006.12611 [astro-ph.HE]].

\bibitem{Zhidenko:2003wq}
A.~Zhidenko,
Class. Quant. Grav. \textbf{21}, 273-280 (2004)
[arXiv:gr-qc/0307012 [gr-qc]].

\bibitem{Cardoso:2003sw}
V.~Cardoso and J.~P.~S.~Lemos,
Phys. Rev. D \textbf{67}, 084020 (2003)
[arXiv:gr-qc/0301078 [gr-qc]].

\bibitem{Tattersall:2018axd}
O.~J.~Tattersall,
Phys. Rev. D \textbf{98}, no.10, 104013 (2018)
[arXiv:1808.10758 [gr-qc]].

\bibitem{Aragon:2020tvq}
A.~Arag\'on, P.~A.~Gonz\'alez, E.~Papantonopoulos and Y.~V\'asquez,
JHEP \textbf{08}, 120 (2020)
[arXiv:2004.09386 [gr-qc]].

\bibitem{Hintz:2021vfl}
P.~Hintz and Y.~Xie,
J. Math. Phys. \textbf{63}, no.1, 011509 (2022)
[arXiv:2105.02347 [gr-qc]].

\bibitem{Churilova:2021nnc}
M.~S.~Churilova, R.~A.~Konoplya and A.~Zhidenko,
Phys. Rev. D \textbf{105}, no.8, 084003 (2022)
[arXiv:2108.04858 [gr-qc]].

\bibitem{Roussev:2022hpm}
A.~Roussev,
Gen. Rel. Grav. \textbf{54}, no.8, 80 (2022)

\bibitem{Konoplya:2007zx}
R.~A.~Konoplya and A.~Zhidenko,
Phys. Rev. D \textbf{76}, no.8, 084018 (2007)
[erratum: Phys. Rev. D \textbf{90}, no.2, 029901 (2014)]
[arXiv:0707.1890 [hep-th]].

\bibitem{Yoshida:2003zz}
S.~Yoshida and T.~Futamase,
Phys. Rev. D \textbf{69}, 064025 (2004)
[arXiv:gr-qc/0308077 [gr-qc]].

\bibitem{Konoplya:2004uk}
R.~A.~Konoplya and A.~Zhidenko,
JHEP \textbf{06}, 037 (2004)
[arXiv:hep-th/0402080 [hep-th]].

\bibitem{Dyatlov:2011jd}
S.~Dyatlov,
Annales Henri Poincare \textbf{13}, 1101-1166 (2012)
[arXiv:1101.1260 [math.AP]].

\bibitem{Konoplya:2003dd}
R.~A.~Konoplya,
Phys. Rev. D \textbf{68}, 124017 (2003)
[arXiv:hep-th/0309030 [hep-th]].

\bibitem{Molina:2003ff}
C.~Molina,
Phys. Rev. D \textbf{68}, 064007 (2003)
[arXiv:gr-qc/0304053 [gr-qc]].

\bibitem{Fernando:2015fha}
S.~Fernando,
Int. J. Mod. Phys. D \textbf{24}, no.14, 1550104 (2015)
[arXiv:1508.03581 [gr-qc]].

\bibitem{Natario:2004jd}
J.~Natario and R.~Schiappa,
Adv. Theor. Math. Phys. \textbf{8}, no.6, 1001-1131 (2004)
[arXiv:hep-th/0411267 [hep-th]].

\bibitem{Konoplya:2008au}
R.~A.~Konoplya and A.~Zhidenko,
Phys. Rev. Lett. \textbf{103}, 161101 (2009)
[arXiv:0809.2822 [hep-th]].

\bibitem{Konoplya:2017ymp}
R.~A.~Konoplya and A.~Zhidenko,
Phys. Rev. D \textbf{95}, no.10, 104005 (2017)
[arXiv:1701.01652 [hep-th]].

\bibitem{Cuyubamba:2016cug}
M.~A.~Cuyubamba, R.~A.~Konoplya and A.~Zhidenko,
Phys. Rev. D \textbf{93}, no.10, 104053 (2016)
[arXiv:1604.03604 [gr-qc]].

\bibitem{Konoplya:2013sba}
R.~A.~Konoplya and A.~Zhidenko,
Phys. Rev. D \textbf{89}, no.2, 024011 (2014)
[arXiv:1309.7667 [hep-th]].

\bibitem{Cardoso:2017soq}
V.~Cardoso, J.~L.~Costa, K.~Destounis, P.~Hintz and A.~Jansen,
Phys. Rev. Lett. \textbf{120}, no.3, 031103 (2018)
[arXiv:1711.10502 [gr-qc]].

\bibitem{Dias:2018ynt}
O.~J.~C.~Dias, F.~C.~Eperon, H.~S.~Reall and J.~E.~Santos,
Phys. Rev. D \textbf{97}, no.10, 104060 (2018)
[arXiv:1801.09694 [gr-qc]].

\bibitem{Dias:2018ufh}
O.~J.~C.~Dias, H.~S.~Reall and J.~E.~Santos,
Class. Quant. Grav. \textbf{36}, no.4, 045005 (2019)
[arXiv:1808.04832 [gr-qc]].

\bibitem{Hod:2020ktb}
S.~Hod,
Int. J. Mod. Phys. D \textbf{29}, no.14, 2042003 (2020)
[arXiv:2012.01449 [gr-qc]].

\bibitem{Liu:2019lon}
H.~Liu, Z.~Tang, K.~Destounis, B.~Wang, E.~Papantonopoulos and H.~Zhang,
JHEP \textbf{03}, 187 (2019)
[arXiv:1902.01865 [gr-qc]].

\bibitem{Price:1971fb}
R.~H.~Price,
Phys. Rev. D \textbf{5}, 2419-2438 (1972).

\bibitem{Brady:1996za}
P.~R.~Brady, C.~M.~Chambers, W.~Krivan and P.~Laguna,
Phys. Rev. D \textbf{55}, 7538-7545 (1997)
[arXiv:gr-qc/9611056 [gr-qc]].

\bibitem{Brady:1999wd}
P.~R.~Brady, C.~M.~Chambers, W.~G.~Laarakkers and E.~Poisson,
Phys. Rev. D \textbf{60}, 064003 (1999)
[arXiv:gr-qc/9902010 [gr-qc]].

\bibitem{Molina:2003dc}
C.~Molina, D.~Giugno, E.~Abdalla and A.~Saa,
Phys. Rev. D \textbf{69}, 104013 (2004)
[arXiv:gr-qc/0309079 [gr-qc]].

\bibitem{Lopez-Ortega:2006aal}
A.~Lopez-Ortega,
Gen. Rel. Grav. \textbf{38}, 1565-1591 (2006)
[arXiv:gr-qc/0605027 [gr-qc]].

\bibitem{Jansen:2017oag}
A.~Jansen,
Eur. Phys. J. Plus \textbf{132} (2017) no.12, 546
[arXiv:1709.09178 [gr-qc]].

\bibitem{Leaver:1985ax}
E.~W.~Leaver,
Proc. Roy. Soc. Lond. A \textbf{402}, 285-298 (1985).

\bibitem{Moss:2001ga}
I.~G.~Moss and J.~P.~Norman,
Class. Quant. Grav. \textbf{19}, 2323-2332 (2002)
[arXiv:gr-qc/0201016 [gr-qc]].

\bibitem{Nollert:1993zz}
H.~P.~Nollert,
Phys. Rev. D \textbf{47}, 5253-5258 (1993).

\bibitem{Cardoso:2006bv}
V.~Cardoso and M.~Cavaglia,
Phys. Rev. D \textbf{74}, 024027 (2006)
[arXiv:gr-qc/0604101 [gr-qc]].

\bibitem{Fortuna:2020obg}
S.~Fortuna and I.~Vega,
[arXiv:2003.06232 [gr-qc]].

\bibitem{Gundlach:1993tp}
C.~Gundlach, R.~H.~Price and J.~Pullin,
Phys. Rev. D \textbf{49}, 883-889 (1994)
[arXiv:gr-qc/9307009 [gr-qc]].

\bibitem{Konoplya:2011qq}
R.~A.~Konoplya and A.~Zhidenko,
Rev. Mod. Phys. \textbf{83}, 793-836 (2011)
[arXiv:1102.4014 [gr-qc]].

\bibitem{Chandrasekhar} S. Chandrasekhar, Proc. Roy. Soc. Lond. A 392 (1984) 1.

\bibitem{Konoplya:2022hll}
R.~A.~Konoplya, A.~F.~Zinhailo, J.~Kunz, Z.~Stuchlik and A.~Zhidenko,
[arXiv:2206.14714 [gr-qc]].

\bibitem{Konoplya:2022pbc}
R.~A.~Konoplya and A.~Zhidenko,
[arXiv:2209.00679 [gr-qc]].

\end{thebibliography}
\end{document}